\documentclass[aps,prd,preprint,superscriptaddress,tightenlines,nofootinbib]{revtex4}

\usepackage{graphicx}
\usepackage{dcolumn}
\usepackage{bm}

\newcommand{\etal}{\textit{et al.}}

\newcommand{\Hszer}{\ensuremath{H^2_0(\qsq)}}
\newcommand{\Hpls}{\ensuremath{H_+(\qsq)}}
\newcommand{\Hmin}{\ensuremath{H_-(\qsq)}}
\newcommand{\Hzer}{\ensuremath{H_0(\qsq)}}
\newcommand{\hzer}{\ensuremath{h_0(\qsq)}}

\newcommand{\hDFzer}{\ensuremath{h_0^{(d,f)}(\qsq)}}
\newcommand{\Hint}{\ensuremath{\hzer\,\Hzer}}

\newcommand{\HDFint}{\ensuremath{\hDFzer\,\Hzer}}

\newcommand{\krzb}{\ensuremath{\overline{K}^{*0}}}
\newcommand{\krzmndk}{\ensuremath{D^+ \rightarrow \krzb e^+ \nu_e}}
\newcommand{\krzmundk}{\ensuremath{D^+ \rightarrow \krzb \mu^+ \nu_\mu}}
\newcommand{\krzlndk}{\ensuremath{D^+ \rightarrow \krzb \ell^+ \nu_\ell}}

\newcommand{\kpimndk}{\ensuremath{D^+ \rightarrow K^- \pi^+ e^+ \nu_e }}

\newcommand{\kpilndk}{\ensuremath{D^+ \rightarrow K^- \pi^+ \ell^+ \nu_\ell}}

\newcommand{\gevcsq}{\ensuremath{\textrm{GeV}/c^2}}

\newcommand{\gevcqd}{\ensuremath{\textrm{GeV}^2/c^4}}

\newcommand{\thv}{\ensuremath{\theta_{V}}}
\newcommand{\thl}{\ensuremath{\theta_\ell}}
\newcommand{\costhv}{\ensuremath{\cos\thv}}

\newcommand{\sinthv}{\ensuremath{\sin\thv}}
\newcommand{\costhl}{\ensuremath{\cos\thl}}

\newcommand{\sinthl}{\ensuremath{\sin\thl}}
\newcommand{\sinthlsq}{\ensuremath{\sin^2\thl}}
\newcommand{\qsq}{\ensuremath{q^2}}

\newcommand{\bw}{\ensuremath{BW}}
\newcommand{\mkpi}{\ensuremath{m_{K\pi}}}

\newcommand{\mysection}[1]{\section{#1}}

\newcounter{saveeqn}%

\begin{document}

\preprint{CLNS 06/1960}       
\preprint{CLEO 06-09}         

\title{Model Independent Measurement of Form Factors in the Decay \kpimndk{}}

\author{M.~R.~Shepherd}
\affiliation{Indiana University, Bloomington, Indiana 47405 }
\author{D.~Besson}
\affiliation{University of Kansas, Lawrence, Kansas 66045}
\author{T.~K.~Pedlar}
\affiliation{Luther College, Decorah, Iowa 52101}
\author{D.~Cronin-Hennessy}
\author{K.~Y.~Gao}
\author{D.~T.~Gong}
\author{J.~Hietala}
\author{Y.~Kubota}
\author{T.~Klein}
\author{B.~W.~Lang}
\author{R.~Poling}
\author{A.~W.~Scott}
\author{A.~Smith}
\author{P.~Zweber}
\affiliation{University of Minnesota, Minneapolis, Minnesota 55455}
% ...
\collaboration{CLEO Collaboration} 
\noaffiliation

\date{June 1, 2006}

\begin{abstract}
We present model independent measurements of the helicity basis form factors
in the decay \kpimndk{}
obtained from about 2\,800 decays reconstructed from a 281 pb$^{-1}$ data
sample collected at the $\psi(3770)$ center-of-mass energy
with the CLEO-c detector.
We confirm the existence of a previously observed spin-zero $K^-\pi^+$
component interfering with the \krzb{} amplitude.
We see no evidence for additional $d$- or $f$-wave contributions. 
\end{abstract}

\maketitle

\mysection{INTRODUCTION}
Exclusive semileptonic decays are excellent probes of charm decay dynamics
since strong interaction effects only enter through the current coupling
the parent $D$ meson with the final state hadronic system~\cite{bigi}.
This current is generally expressed in terms of form factors that depend on
\qsq{}, the squared invariant mass of the virtual $W^+$ materializing as a
positron-neutrino pair. The specific \qsq{} dependence is an ansatz in most of
the models, although the ultimate goal is to derive both normalization
and \qsq{} dependence from first principles.
Various theoretical approaches have been pursued to study semileptonic
form factors, such as quark models~\cite{quark},
QCD sum rules~\cite{ball}, lattice QCD~\cite{lqcd}, and the parameterization
of the form factor \qsq{} dependence based on effective poles~\cite{slovenia}
constrained by heavy quark and soft collinear effective theories.
Measurements of both the branching fraction~\cite{relbr} and
the \qsq{} dependence of the form factors~\cite{oldformfactor}
have been made for the \krzlndk{} decays using specific
parameterizations~\cite{oldformfactor}. 
Recently, Ref.~\cite{anomaly} reported a first observation of an additional
component besides the dominant \krzlndk{} in the \kpilndk{} decay process.
In the additional component, $K^-$ and $\pi^+$ are in a relative $s$-wave,
which has revealed an interesting connection
between semileptonic decays and light quark physics.  According to 
Ref.~\cite{formfactor},  $2.4 \pm 0.7\%$ of the decays in
the mass range $0.8~\gevcsq{} < \mkpi{} < 1.0~\gevcsq{}$ are due to
this $s$-wave component, where \mkpi{} is the $K^- \pi^+$ mass.

Using a technique developed by
FOCUS~\cite{helicity-focus}, we present non-parametric measurements
of the \qsq{} dependence of the helicity basis form factors that describe 
an amplitude for the $K^- \pi^+$ system to be in any one of its possible
angular momentum states. This is done by projecting out the helicity 
form factors directly from data without the use of fitting functions.
Our results will allow theorists to directly compare their models with 
the data free from parameterization.
We also confirm the existence of the $s$-wave
component, study its form factor for the first time, and limit
the strength of possible $d$- and $f$-wave contributions.

There are several good reasons to perform this
analysis using the CLEO-c data: The \qsq{} resolution in \mbox{CLEO-c} is
an order of magnitude better than in FOCUS.  In addition,
the \krzmndk{} process is simpler than \krzmundk{}, since there
is a negligible probability for the much less massive $e^+$ to be left-handed
which eliminates one of the form factors.
The additional form factor describing the left-handed $\ell^+$ coupling
has an angular distribution so similar to $H_0$ in Eq.~(\ref{amp1})
that it degrades the measurement of $H_0$ by a factor of three.

The amplitude $\mathcal M$ for the semileptonic decay \kpimndk{} is described
by five kinematic quantities: 
\qsq{}, \mkpi{}, the angle between the $\pi$ and the $D$
direction in the $K^- \pi^+$ rest frame ($\thv$),
the angle between the $\nu_e$ and the $D$ direction in the $e^+ \nu_e$
rest frame ($\thl$), 
and the acoplanarity angle between the two decay planes ($\chi$).
Following Ref.~\cite{formfactor}, we can express the matrix element
for the decay \kpimndk{} in the vicinity of the \krzb{} mass in terms of
the three helicity amplitudes (\Hpls{},\Hmin{}, and \Hzer{})
describing the pseudoscalar to vector hadronic 
current and one form factor (\hzer{}) describing a broad $s$-wave resonance.
The dominant terms in the differential width $|\mathcal{M}|^2$,
integrated over the angle $\chi$ can be expressed as: 
\begin{eqnarray}
\lefteqn{ \int { |\mathcal{M}|^2 d\chi }
          \propto G_F^2 \left| V_{cs} \right|^2
             (q^2 - m_\ell ^2)    } \nonumber \\ 
  & \times & [\,
          ((1 + \costhl) \sinthv)^2 |H_+(q^2)|^2 |\bw|^2  \nonumber \\ 
  & &     + ((1 - \costhl) \sinthv)^2 |H_-(q^2)|^2 |\bw|^2  \label{amp1}\\ 
  & &     + (2 \sinthl \costhv)^2 |H_0(q^2)|^2 |\bw|^2  \nonumber \\ 
  & &     + 8 \sinthlsq \costhv H_0(q^2)h_0(q^2)\,
                {\mathop{\mathrm{Re}}\nolimits}\{\mathcal{A}e^{-i\delta} \bw \}
                \nonumber \\
  & &     + \mathcal{O}(\mathcal{A}^2)
                     \, ] \,.\nonumber 
\end{eqnarray}
The second to last term in Eq.~(\ref{amp1}) represents the interference between
the $s$-wave $K^- \pi^+$ and the \krzb{} amplitude, where $\mathcal{A}$ and
$\delta$ represent the amplitude and the phase of the $s$-wave, respectively.
Since the $s$-wave amplitude could be observed only through interference
with the \krzb{}, FOCUS was
sensitive to the $s$-wave amplitude only in the vicinity of the \krzb{} pole.
They modeled the \mbox{$s$-wave} contribution as a constant complex amplitude,
even though a variation of the amplitude with respect to \mkpi{} is possible.
In this paper, we assume the $\mathcal{A}$ and $\delta$ values obtained
in Ref.~\cite{formfactor} in the study of the \Hint{} interference term.
The \krzb{} amplitude in Eq.~(\ref{amp1}) is represented
as a Breit-Wigner of the form:
\[
\bw = \frac{\sqrt{m_0 \Gamma} \left(\frac{P^*}{P_0^*}\right)}
	   {m_{K\pi}^2 - m_0^2 + i m_0 \Gamma 
	      \left(\frac{P^*}{P_0^*}\right)^3}\, , 
\]
where $P^*$ is the kaon momentum in the $K^- \pi^+$ rest frame and
$P_0^*$ is the kaon momentum in this frame at the resonant mass $m_0$.
The $\chi$ integration significantly simplifies the intensity by eliminating
all interference terms between different helicity states of the virtual $W^+$
with relatively little loss in form factor information. We have dropped
the term which is second order in the small, $s$-wave amplitude $\mathcal{A}$.

The three helicity basis form factors for the \krzmndk{} component are
generally written~\cite{KS} as linear combinations of vector ($V(\qsq{})$)
and axial-vector ($A_{1,2}(\qsq{})$) form factors 
according to Eq.~(\ref{KS}): 
\begin{widetext}
\begin{eqnarray}
H_\pm (\qsq) &=&
   (M_D+\mkpi)A_1(\qsq)\mp 2{M_D \tilde{K}\over M_D+m_{K\pi}}V(\qsq) \label{KS}
\quad \textrm{and} \\
H_0 (\qsq) &=& 
   {1\over 2\mkpi\sqrt{\qsq}}
   \left[ (M^2_D -m^2_{K\pi}-\qsq)(M_D+\mkpi)A_1(\qsq)
   -4{M^2_D \tilde{K}^2\over M_D+\mkpi}A_2(\qsq) \right],  \nonumber
\end{eqnarray}
\end{widetext}
where $M_D$ is the mass of the $D^+$ and $\tilde{K}$ is the momentum
of the $K^- \pi^+$
system in the rest frame of the $D^+$.

In this paper, we use a \emph{projective weighting}
technique~\cite{helicity-focus} to disentangle and directly measure the \qsq{}
dependence of these helicity basis form factors free from parameterization.
We provide information on the form factor products
$H_\pm^2(\qsq)$, $H_0^2(\qsq)$, and $h_0(\qsq) H_0(\qsq)$ in bins
of \qsq{} by projecting out the associated angular factors given
by Eq.~(\ref{amp1}).

\mysection{\label{exp} EXPERIMENTAL AND ANALYSIS DETAILS}
The CLEO-c detector~\cite{detector} consists of a six-layer, low-mass,
stereo-wire drift chamber,
a 47-layer central drift chamber, a ring-imaging \v{C}erenkov detector
(RICH), and a cesium iodide electromagnetic calorimeter inside a superconducting
solenoidal magnet providing a 1.0~T magnetic field. The tracking chambers
and the electromagnetic calorimeter cover 93\% of the full solid angle.
The solid angle coverage for the RICH detector is 80\% of $4\pi$. Identification
of the charged pions and kaons is based on measurements of specific
ionization ($dE/dx$) in the main drift chamber and RICH information.
In positron identification, in addition to $dE/dx$ and RICH information,
the ratio of energy deposited in the electromagnetic calorimeter to
the measured track momentum ($E/p$) is used.

In this paper, we use 281~pb$^{-1}$ of data taken at the $\psi(3770)$
center-of-mass energy with the CLEO-c detector at the Cornell Electron Storage
Ring (CESR), which corresponds to a sample of
0.8 million $D^+D^-$ pair events~\cite{dtag}. 
Monte Carlo (MC) events are generated by \textsc{evtgen}~\cite{evtgen} and the
detector is simulated using a \textsc{geant}-based~\cite{geant} program.
Simulation of final state radiation is handled by \textsc{photos}\cite{photos}. 
Throughout this paper charge-conjugate modes are implied.

We select events where a \kpimndk{} candidate is produced against
a fully reconstructed tagging $D^-$. 
The tagging $D^-$ decays into one of the following six decay modes:
$D^- \rightarrow K^0_S\pi^-$, $D^- \rightarrow K^+\pi^-\pi^-$,
$D^- \rightarrow K^0_S\pi^-\pi^0$, $D^- \rightarrow K^+\pi^-\pi^-\pi^0$,
$D^- \rightarrow K^0_S\pi^-\pi^-\pi^+$, and $D^- \rightarrow K^-K^+\pi^-$.
Multiple $D^-$ candidates per event are allowed.
More details on selecting the tagging $D^-$ candidates as well as identifying
$\pi^0$ and $K^0_S$ candidates are described in Ref.~\cite{dtag}.

\begin{figure}
\includegraphics[width=2.8in]{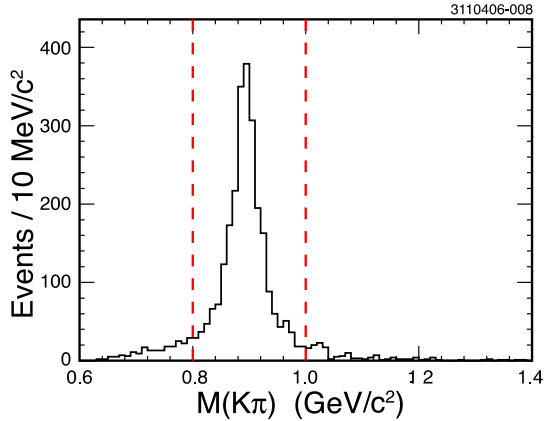}
\caption{ The \mkpi{} distribution for events satisfying our \emph{nominal}
\kpimndk{} selection requirements. Over the full displayed mass range, there
   are 2\,838 events satisfying our nominal \kpimndk{} selection.
   For this analysis, we use a restricted mass range 0.8 $-$ 1.0 \gevcsq{}
   (shown by the vertical dashed lines). In this restricted region, there are
   2\,472 events.  
   \label{signal}}
\end{figure}

The following selection cuts represent our \emph{nominal selection}
criteria.  The semileptonic $D^+$ reconstruction starts by requiring
three well-measured tracks not coming from the tagging $D^-$ decay in the event.
Charged kaons and pions are required to have momenta of at least 50~MeV/$c$
and are
identified using $dE/dx$ and RICH information. Positron candidates are
required to have momenta of at least 200~MeV/$c$, satisfy
$|\cos \theta| < 0.9$, where $\theta$ is the angle between the positron 
and the beam line, and pass a requirement on a 
likelihood variable that combines $E/p$, $dE/dx$, and RICH information.
We obtain 2\,838 \kpimndk{} events.  Finally, we require
$0.8~\gevcsq{} \le \mkpi \le 1.0~\gevcsq{}$ and select 2\,472 events.
The \mkpi{} distribution for these \kpimndk{} candidates is shown in
Fig.~\ref{signal}.
In our analysis, the only particle not reconstructed is the neutrino of
the \kpimndk{} decay.
The neutrino four-momentum can be reconstructed from the energy-momentum balance
in the beam-beam center-of-mass frame.
We assign the neutrino the missing energy ($E_\mathrm{miss}$) which is
the difference between the beam energy ($E_\mathrm{beam}$)
and the sum of the energies of the charged semileptonic daughters.
The neutrino momentum is obtained from direction of the missing momentum
($\hat{p}_\mathrm{miss} = \vec p_\mathrm{miss}/ | \vec p_\mathrm{miss}|$)
defined as: 
\begin{equation}
\vec p_\mathrm{miss} = - \sqrt{E_\mathrm{beam}^2 - m_D^2}
                   \, \hat p_{D^-}
                 - \vec p_{K^-} - \vec p_{\pi^+} - \vec p_{e^+} .
                \label{closure}
\end{equation}
In Eq.~(\ref{closure}),
$\hat p_{D^-}$ is the measured direction of the tagging $D^-$, 
$\vec p$ is the measured momentum of the indicated particle, and
$m_D$ is the known mass of the $D^-$~\cite{pdg}.
The magnitude of the neutrino momentum is set to $E_\mathrm{miss}$.

\begin{figure}
\includegraphics[width=2.8in]{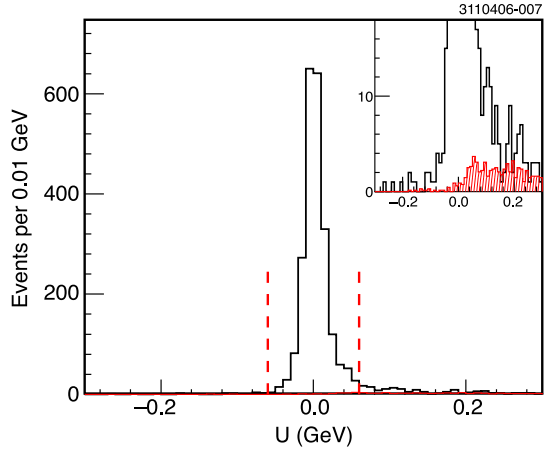}
\caption{The observed energy imbalance $U$ distribution for \kpimndk{} events
    satisfying our \emph{nominal} selection. We show the boundaries for the
    additional $|U| < 0.06$ GeV requirement used for the \emph{tighter}
    selection.
    The inset is the same distribution on a finer scale
    where the background contribution estimated from a Monte Carlo sample
    is more apparent as a hashed histogram.
   \label{Udist}}
\end{figure}

We will compare the \emph{nominal} selection to a \emph{tighter} selection where
we add two further restrictions: $0.846~\gevcsq{} \le \mkpi
\le 0.946~\gevcsq{}$ 
and $|U| < 0.06$ GeV, where 
$U \equiv E_\mathrm{miss} - c|\vec p_\mathrm{miss}|$.
Figure~\ref{Udist} shows the $U$ distribution for the nominally selected sample 
with the background contribution. 
The background shape in Fig.~\ref{Udist} is obtained
using a charm MC set which consists of generic $D\overline{D}$ events.  
We estimate a background level of 3.6\% for the \emph{nominal} selection
and 0.4\% for the \emph{tighter} selection.
We have checked the MC background simulation with a sample of
wrong sign events where we reconstructed the semileptonic candidates
with electrons instead of positrons and obtained consistent results between
the data and the MC samples.

\begin{figure}
\includegraphics[width=2.8in]{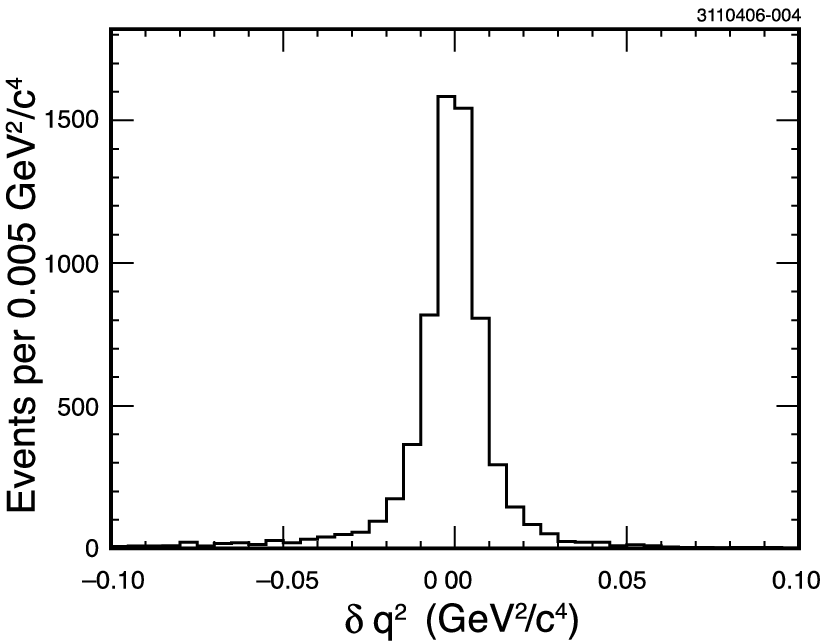}
\caption{The \qsq{} resolution predicted by the generic $D\overline{D}$ MC. 
   \label{qsqres}}
\end{figure}

The \qsq{} resolution predicted by our MC is roughly Gaussian with an r.m.s.
width of 0.02 \gevcqd{}, which is shown in Fig.~\ref{qsqres}.
This is negligible on the scale in which we will bin our data.

Even though we only measure the shape and not absolute rates, as a 
cross-check we have measured the \krzmndk{} branching fraction
assuming an efficiency derived from the generic MC.
We found statistical consistency with the result in Ref.~\cite{victor}.

\mysection{PROJECTION WEIGHTING TECHNIQUE} 
We extract the helicity basis form factors using the projective weighting
technique more fully described in Ref.~\cite{helicity-focus}.  For a given
\qsq{} bin, a weight is assigned to the event depending on its \thv{} and \thl{}
decay angles.  We use 25 joint  $\Delta \costhv \times \Delta \costhl$ angular
bins:  5 evenly spaced bins in \costhv{} times 5 bins in \costhl{}. 

For each $q_i^2$ bin, the angular distribution can be written
as a vector $\vec N_i$ whose components give the number of events
reconstructed in each of the 25 angular bins. 
According to Eq.~(\ref{amp1}), $\vec N_i$ can be written as a linear
combination of $\vec m_{\alpha}$ vectors with coefficients $f_\alpha(q_i^2)$,
\begin{equation}
  \vec N_i = f_+(q_i^2)\,\vec m_+ + f_-(q_i^2)\,\vec m_-
            + f_0(q_i^2)\,\vec m_0 + f_I(q_i^2)\,\vec m_I \, .
\label{series}
\end{equation}
The $\vec m_+$, $\vec m_-$, $\vec m_0$, and $\vec m_I$ vectors are computed
using a full detector simulation weighted by the corresponding
helicity term in Eq.~(\ref{amp1}). 
For example, $\vec m_+$ is computed using a simulation
generated with an arbitrary function for $H_+ (q^2)$ (such as $H_+ (q^2) = 1$) 
and zero for the remaining three form factors. 
The $f_\alpha(q_i^2)$ functions are proportional to the true
$H^2_\alpha(q_i^2)$ along with multiplicative factors such as
$G_F^2 \left| V_{cs} \right|^2 (q^2 - m_\ell ^2)$
and additional corrections accounting for experimental effects
such as acceptance.  

\begin{figure*}[htp]
\includegraphics[height=3.8in]{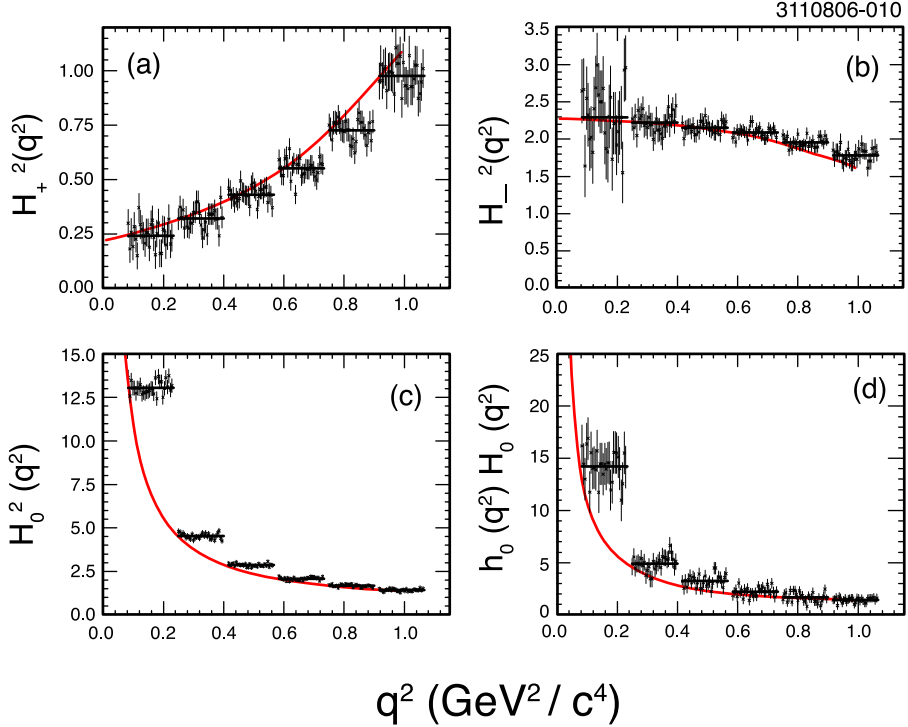}
\caption{Monte Carlo study of the projective weighting technique errors and
   biases.  The reconstructed form factors are
   the points with error bars.  The curves give the helicity form factor
   products assumed in the simulations.  Thirty independent simulations were
   used for each \qsq{} bin. The averages of the simulated results are
   given by the horizontal lines.  Their actual ordinate is at the left of
   the averaging line. The yield in these MC samples are six times the yield
   in the data.  The plots are: 
   (a)~$H_+^2(\qsq)$,
   (b)~$H_-^2(\qsq)$,
   (c)~$H_0^2(\qsq)$, and
   (d)~$h_0(\qsq) H_0(\qsq)$.
\label{mmc}} 
\end{figure*}

Ref.~\cite{helicity-focus} shows how Eq.~(\ref{series}) can be solved
for the four form factor products $H_+^2(\qsq)$, $H_-^2(\qsq)$, $H_0^2(\qsq)$,
and $h_0(\qsq) H_0(\qsq)$ by making four weighted $q^2$ histograms. 
The weights are directly constructed from the four $\vec m_\alpha$ vectors.
We have tested the method using thirty independent Monte Carlo samples.
Figure~\ref{mmc} demonstrates that the projective
weighting technique returns realistic errors with no significant bias.

\begin{figure}[htp]
\includegraphics[height=2.5in]{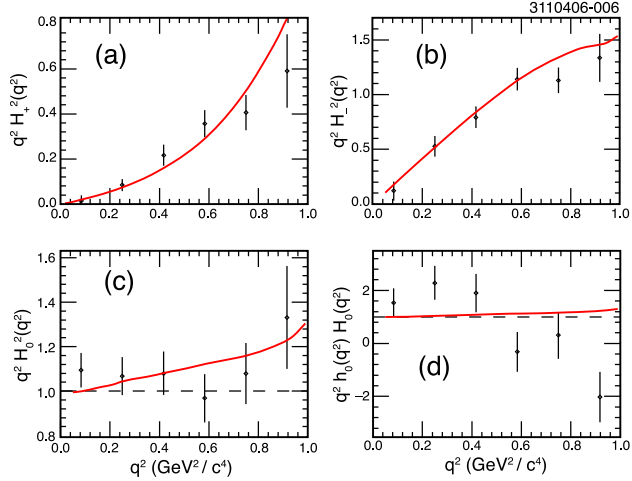}
\caption{
Measured form factor products for six \qsq{} bins.  The data are points
with error bars representing only the statistical uncertainties.
The solid curves are obtained using the form factor representation
and the parameters reported in Ref.~\cite{formfactor},
namely: $V(0)/A_1(0) = 1.505$, $A_2(0)/A_1(0) = 0.875$, and
$s$-wave parameters $\mathcal{A} = 0.33$ and $\delta = 39^\circ$. 
The histogram plots are:
(a)~$\qsq{} H_+^2(\qsq)$,
(b)~$\qsq{} H_-^2(\qsq)$,
(c)~$\qsq{} H_0^2(\qsq)$, and
(d)~$\qsq{} h_0(\qsq) H_0(\qsq)$.
\label{sixnormZ}}
\end{figure}

\mysection{\label{results} RESULTS}
Figure~\ref{sixnormZ} shows the four form factor products 
multiplied by \qsq{} obtained for data using the nominal selection criteria.
In this Figure, the background estimated from the generic charm MC simulation
is subtracted. 
For the nominal selection we estimate that 3.6\% of the events are background.
The form factors are normalized by scaling the four weighted histograms
($H_\pm^2(\qsq)$, $H_0^2(\qsq)$, and $h_0(\qsq)H_0(\qsq)$)
by a common factor so that $\qsq{} \Hszer{} = 1$ as
$\qsq{} \rightarrow 0$.\footnote{The details of the normalization procedure
are as follows: The solid curve in Fig.~\ref{sixnormZ} (c) represents the
parameterized model for the form factor $\qsq{} H_0^2(\qsq{})$ presented
in Ref.~\cite{formfactor}.
This curve is normalized to 1 at $\qsq{} \rightarrow 0$. Then our data points
in Fig.~\ref{sixnormZ} (c) are scaled to the curve by a factor obtained
from a $\chi^2$ fit. The data points in the other Fig.~\ref{sixnormZ}
histograms are all scaled by the same factor.}
Because of our excellent \qsq{} resolution, there is negligible correlation
among the six \qsq{} bins for a given form factor product, but the relative
correlations between different form factor products in the same \qsq{}
bin can be as high as 36\%. The \Hint{} form factor in Fig.~\ref{sixnormZ} (d)
is measured through the interference term of Eq.~(\ref{amp1}),
which is proportional to 
${\mathop{\mathrm{Re}}\nolimits}\{\mathcal{A}e^{-i\delta} \bw \}$.
Averaging over the full \mkpi{} mass range, 
${\mathop{\mathrm{Re}}\nolimits}\{\mathcal{A}e^{-i\delta} \bw \}$
is proportional to $-\mathcal{A} \sin\delta$.
The interference term is then measured by a fourth projective weight and
divided by the FOCUS $-\mathcal{A} \sin\delta$ value~\cite{formfactor}
to obtain \Hint{}.
The measurement on \Hint{} includes
a systematic uncertainty due to uncertainties in $\mathcal{A}\sin\delta$,
which we discuss below.

\begin{figure}[htp]
\includegraphics[height=2.5in]{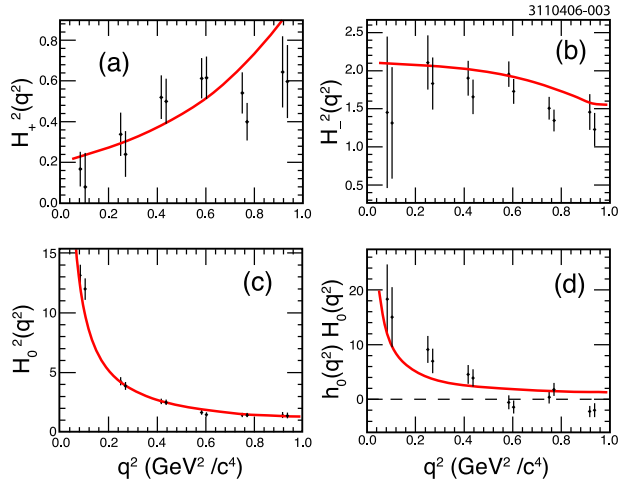}
\caption{
Measured form factor products for six \qsq{} bins.  We show the results 
obtained with the nominal selection criteria (1st point in each bin) and 
the tighter selection criteria (2nd point in each bin). The error bars 
represent statistical uncertainties only. The solid curves are obtained 
using the form factor representation and the parameters reported in
Ref.~\cite{formfactor}.
The histogram plots are:
(a)~$H_+^2(\qsq)$,
(b)~$H_-^2(\qsq)$,
(c)~$H_0^2(\qsq)$, and
(d)~$h_0(\qsq) H_0(\qsq)$.
\label{nom-tight}}
\end{figure}

Figure~\ref{nom-tight} shows the helicity amplitudes without \qsq{}
weighting, thus emphasizing the low \qsq{} region.  It includes results
obtained with the nominal selection criteria as well as the more 
restrictive selection criteria.  The background fraction in the
tightly selected sample is only 0.4\% of the event sample and
its background subtraction is negligible. The consistency of the two results
shows that the systematic error due to background subtraction is small. 

We have considered several sources of systematic uncertainties. Even though
we have very little background in our data set, the background subtraction
still constitutes the primary source of systematic error. 
For this uncertainty, we assign a conservative value 
by reducing the level of the background being subtracted by
a factor of two and comparing these form factor products with the results
based on the full background subtraction. For $H_+^2(\qsq)$,
the background systematic
uncertainty is estimated to be 16.5\% for the first \qsq{} bin and
less than 4.8\% for higher bins. For $H_-^2(\qsq)$, the background systematic
uncertainty is estimated to be 41.3\% for the first \qsq{} bin and
less than 8.1\% for higher bins. For $H_0^2(\qsq)$, the background systematic
uncertainty is estimated to be less than 2.1\%; and for $h_0 H_0$,
the background systematic uncertainty is estimated as less than 18.0\%. 
The systematic uncertainty due to track reconstruction and particle
identification is rather low since we are reporting on
form factor shapes rather than absolute normalization.
The systematic uncertainty from these sources
is estimated as less than 1.9\% for all of the form factor products.
Lastly, we assess an additional scale error of 13.4\% for the $h_0 H_0$
form factor product due to the uncertainties in the $\mathcal{A}$
and $\delta$ values reported in Ref.~\cite{formfactor}.
Our total systematic uncertainty is the quadrature sum of these systematic
contributions and found to be small compared to the statistical
uncertainty. In all cases except $h_0 H_0$, the systematic uncertainty is
dominated by the background systematic uncertainty.
Table~\ref{summary} gives the \qsq{} range for each bin, along with the form
factor products, their statistical uncertainty (first error) and
the estimated systematic uncertainty (second error).

\begin{table*}[htp]
\caption{Summary of form factor product results for six \qsq{} bins. 
Each form factor product is averaged over the indicated \qsq{} range.
The first and second errors are statistical and systematical uncertainties,
respectively. The numbers are normalized using the condition:
$\qsq{} \Hszer{} = 1$ as $\qsq{} \rightarrow 0$.}
\begin{ruledtabular}
\begin{tabular}{ccccc}
\qsq{} range $(\gevcqd{})$ & $H_+^2$ & $H_-^2$ & $H_0^2$  & $h_0 H_0$  \\ 
\hline
0.000$\ -\ $0.167 & 0.16 $\pm$ 0.12 $\pm$ 0.03 & 1.19 $\pm$ 0.87 $\pm$ 0.49 & 
       13.17 $\pm$ 0.91 $\pm$ 0.13 & 13.60 $\pm$ 4.72 $\pm$ 1.86  \\ 
0.167$\ -\ $0.334 & 0.32 $\pm$ 0.10 $\pm$ 0.02 & 1.99 $\pm$ 0.34 $\pm$ 0.16 &
        4.30 $\pm$ 0.34 $\pm$ 0.04 & 6.73 $\pm$ 1.84 $\pm$ 0.92  \\ 
0.334$\ -\ $0.501 & 0.52 $\pm$ 0.11 $\pm$ 0.01 & 1.88 $\pm$ 0.23 $\pm$ 0.08 &
        2.59 $\pm$ 0.23 $\pm$ 0.04 & 3.34 $\pm$ 1.25 $\pm$ 0.46  \\ 
0.501$\ -\ $0.668 & 0.61 $\pm$ 0.10 $\pm$ 0.01 & 1.93 $\pm$ 0.17 $\pm$ 0.08 &
        1.69 $\pm$ 0.18 $\pm$ 0.01 & -0.36 $\pm$ 0.94 $\pm$ 0.08  \\ 
0.668$\ -\ $0.835 & 0.55 $\pm$ 0.10 $\pm$ 0.01 & 1.53 $\pm$ 0.15 $\pm$ 0.03 &
        1.43 $\pm$ 0.18 $\pm$ 0.03 & 0.35 $\pm$ 0.87 $\pm$ 0.06   \\ 
0.835$\ -\ $1.000 & 0.65 $\pm$ 0.18 $\pm$ 0.01 & 1.46 $\pm$ 0.23 $\pm$ 0.03 &
        1.47 $\pm$ 0.25 $\pm$ 0.02 & -1.78 $\pm$ 0.84 $\pm$ 0.24  \\ 
\end{tabular}
\end{ruledtabular}
\label{summary}
\end{table*}

\begin{figure}[tbph!]
\includegraphics[width=3.25in]{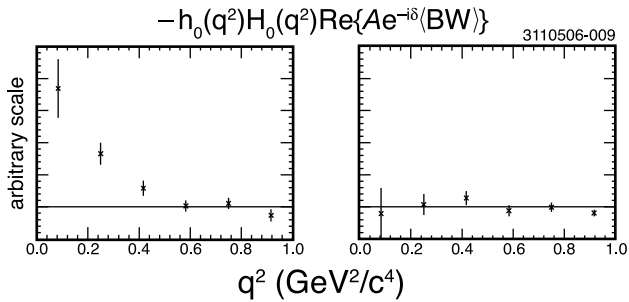}
\caption{The $s$-wave interference term for the events
(a) below the \krzb{} pole and (b) above the pole.
The interference term depends on the $s$-wave phase relative to the 
average phase of the Breit-Wigner in the corresponding plot.
All of the \costhv{} interference observed by FOCUS~\cite{anomaly}
was also below the \krzb{} pole. 
\label{split}}
\end{figure}

Figures~\ref{sixnormZ}~(d) and \ref{nom-tight}~(d) show that
the $h_0(\qsq) H_0(\qsq)$ form factor product is
in rough agreement with the model from Ref.~\cite{formfactor}. 
We are also consistent with the FOCUS $s$-wave phase $\delta = 39 \pm 5^\circ$.
Our consistency check is the comparison of the $s$-wave interference term for
events with \mkpi{} below and above the \krzb{} pole as shown
in Fig.~\ref{split}.
The interference between the $s$-wave and the Breit-Wigner is proportional to 
$\cos(\delta - \beta_\pm)$ where
$\beta_-$ and $\beta_+$ are the average Breit-Wigner phases below and 
above the \krzb{} pole, respectively.
In the mass region where $m(\krzb) < \mkpi < 1.0$ GeV, $\beta_+$ is 
$-59^\circ$, which is nearly orthogonal to the FOCUS $s$-wave phase.
If our phase is consistent with the FOCUS phase, the effective
$\Hint {\mathop{\mathrm{Re}}\nolimits} \{ 
\mathcal{A} e^{ - i\delta } \left\langle {BW} \right\rangle \}$ should
disappear above the \krzb{} pole as it does in Fig.~\ref{split}~(b).
The sum of the first three bins of the histogram of Fig.~\ref{split}~(a)
differs from zero by 5.9 standard deviations.
This confirms the earlier observation~\cite{anomaly} of an $s$-wave interference
with the FOCUS phase, but does not provide an independent measurement of
that phase.

It is of interest to search for the possible
existence of additional non-resonant amplitudes of higher angular momentum. 
It is fairly simple to extend Eq.~(\ref{amp1}) to account for potential
$d$-wave or $f$-wave interference with the \krzb{} Breit-Wigner.
We search specifically for a possible zero helicity $d$-wave or $f$-wave
piece that interferes with the zero helicity \krzb{} contribution. One expects
that potential $h^{(d)}_0(q^2)$ and $h^{(f)}_0(q^2)$ form factors would
peak as $1/\sqrt{\qsq}$ near $\qsq{} \rightarrow 0$ as is the case for the
other zero helicity contributions \Hzer{} and \hzer{}. If so, the zero helicity
contributions should be much larger than potential $d$- or $f$-wave $\pm 1$
helicity contributions. 
The $d$- and $f$-wave interference terms are the same as the $s$-wave interference
term apart from additional Wigner $d$-matrices. These describe anisotropy
in the $d$- and $f$-wave decays. The additional factors are $d^2_{0,0}(\thv)$
for the $d$-wave and $d^3_{0,0}(\thv)$ for the $f$-wave. 
Hence, the $d$-wave weights are based on an fifth term
of the form 
$4\,\sinthlsq \cos \thv\,(3\,\cos^2 \thv - 1)\,H_0(q^2)\,h^{(d)}_0(q^2)\,
{\mathop{\mathrm{Re}}\nolimits}\{\mathcal{A}_d\,e^{-i\delta_d} \bw\}$
in Eq.~(\ref{amp1}).
The $f$-wave weights are based on an fifth term of the form
$4\,\sinthlsq \cos \thv\,(5\,\cos^3 \thv - 3\cos\thv)\,H_0(q^2)\,h^{(f)}_0(q^2)\,
{\mathop{\mathrm{Re}}\nolimits}\{\mathcal{A}_f\,e^{-i\delta_f} \bw \}$. 
Averaging over the Breit-Wigner, the interference should be proportional to
$\mathcal{A}_{d,f}\,\sin{\delta_{d,f}}\,\HDFint{}$.

\begin{figure*}[htp]
\includegraphics[height=1.8in]{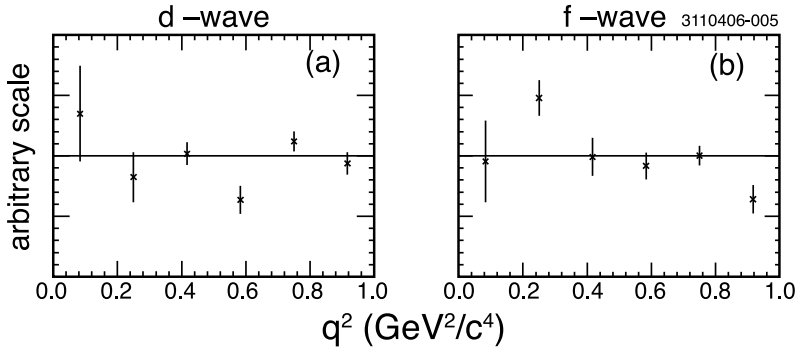}
\caption{Search for (a) $d$-wave and (b) $f$-wave interference effects,
$\mathcal{A}_{d,f}\,\sin{\delta_{d,f}}\,\HDFint{}$, as described in the text. 
\label{wavelimits}}
\end{figure*}
Figure~\ref{wavelimits} shows the 
$\mathcal{A}_{d,f}\,\sin{\delta_{d,f}}\,\HDFint{}$ 
obtained with this technique.
It is apparent from Fig.~\ref{wavelimits} that we have no compelling evidence
for either a $d$-wave or an $f$-wave component. Unfortunately, we cannot set
a particularly stringent limit. Under the assumption
$h^{(d,f)}_0(\qsq{}) = \Hzer{}$, used in Ref.~\cite{formfactor}, 
we perform a $\chi^2$ fit of Fig.~\ref{wavelimits} to the form
$\mathcal{A}_{d,f}\,\sin{\delta_{d,f}}\,H^2_0(\qsq)$. 
The results of these fit were 
$\mathcal{A}_d\,\sin{\delta_d} = -0.07 \pm 0.20$ and
$\mathcal{A}_f\,\sin{\delta_f} = 0.17 \pm 0.18$. For comparison, the value for
$\mathcal{A}\,\sin{\delta}$ for the $s$-wave contribution according to
Ref.~\cite{formfactor} is $0.21 \pm 0.028$.

\mysection{SUMMARY}
We present a non-parametric analysis of the helicity basis form factors that
control the kinematics of the \kpimndk{} decays. We use a projective weighting
technique that allows one to determine the helicity form factor products
by weighted histograms rather than likelihood based fits.
The resulting helicity form factors have very little background
and the background subtraction procedure is well understood.
We confirm the existence of an $s$-wave $K^- \pi^+$ component to
the \kpimndk{} decay and have studied the \qsq{} dependence of its form factor.
Finally, we have searched for possible $d$-wave or $f$-wave non-resonant
interference contributions to \kpimndk{}. Although we have no statistically
significant evidence for $f$-wave or $d$-wave interference, we are unable
to limit these terms to appreciably less than the established $s$-wave
interference.


We gratefully acknowledge the effort of the CESR staff 
in providing us with excellent luminosity and running conditions. 
D.~Cronin-Hennessy and A.~Ryd thank the A.P.~Sloan Foundation. 
This work was supported by the National Science Foundation,
the U.S. Department of Energy, and 
the Natural Sciences and Engineering Research Council of Canada.

\end{document}